\documentstyle[pra,aps,amsfonts,graphicx]{revtex}
\begin{document}
\draft
\twocolumn[\hsize\textwidth\columnwidth\hsize\csname 
@twocolumnfalse\endcsname
\title{Image formation by manipulation of the entangled angular spectrum}
\author{D. P. Caetano$^{*}$ and P. H. Souto Ribeiro}
\address{Instituto de F\'{\i}sica, Universidade Federal do Rio de Janeiro, Caixa Postal 68528, Rio de Janeiro, RJ 21941-972, Brazil} 
\date{\today}
\maketitle

\begin{abstract}
We demonstrate theoretical and experimentally how it is possible to manipulate an entangled angular spectrum of twin beams, in order to reconstruct correlated images with coincidence detection. The entangled angular spectrum comes from the pump and the image is obtained only if signal and idler are properly treated.
\end{abstract}

]

\section{Introduction}

Image formation and spatial patterns in optical parametric processes have been subject of interest in the last years either in classical\cite{cimages} or in quantum optics\cite{qimages}. In the process of spontaneous parametric down-conversion, twin photons are generated with strong correlation in the transverse momentum, as a consequence of the entanglement in the wave vector. Many experiments have explored the correlations in this degree of freedom to observe conditional interference patterns\cite{interfconditional}, quantum imaging\cite{shih2,barbosa,spectrumtrans,imagepol}, and quantum correlations between optical modes in the Laguerre-Gaussian\cite{OAM} and Hermite-Gaussian\cite{HGmodes} basis.

Image formation in spontaneous parametric down-conversion has been demonstrated by placing diffracting masks and lenses either in the twin beams\cite{imagetwin} or in the pump beam\cite{imagepump} and detecting the twin beams in coincidence. In Ref. \cite{spectrumtrans} it was demonstrated that the transverse properties of the pump beam are connected with the transverse correlations of the twin beams. In some cases, the coincidence transverse pattern exactly reproduces the transverse intensity pattern of the pump beam.

In this work, we explore the transfer of the angular  spectrum from the pump to the signal-idler entangled spectrum\cite{spectrumtrans}, in order to investigate, theoretical and experimentally, image formation in the coincidence transverse pattern, when a diffracting mask is placed in the pump and the imaging lenses are placed in the signal and idler beams.

\section{Theory}
Let us consider the situation sketched in Fig. 1. A laser beam (pump) illuminates an object $\cal O$ and
pumps a nonlinear crystal placed at a distance $z_1$ from the object. By spontaneous parametric down-conversion, signal and idler twin beams are emitted
collinearly with the pump. They are directed to a lens L placed at a distance $z_2$ from the crystal. An optical window (M) reflects the violet radiation and transmits the infrared radiation separating the pump from the twin beams. The twin beams are detected in coincidence after a beam-splitter (BS) by detectors D1 and D2. Both detectors are placed at a distance $z$ from the crystal. The pump is detected by the detector D3, also at a distance $z$ from the crystal. The lens L is such that it generates the image of the object $\cal O$ at the image plane, corresponding to the detection plane.

The state of the field generated by the spontaneous parametric down-conversion in the monochromatic and paraxial approximations can be written as\cite{spectrumtrans}

\begin{equation}
\label{eq1}
|\psi\rangle = \alpha|vac\rangle + \beta\int d{\bf q}_i \int d{\bf q}_s \, {\cal V}_{z_2}({\bf q}_i+{\bf q}_s)|1,{\bf q}_i\rangle |1,{\bf q}_s\rangle,
\end{equation}
where $|1,{\bf q}_s\rangle$ and $|1,{\bf q}_i\rangle$ are 1 photon Fock states with transverse wave vectors ${\bf q}_s$ and ${\bf q}_i$, corresponding to the down-converted modes signal and idler, respectively. $|vac\rangle$ is the vacuum state. ${\cal V}_{z_2}({\bf q}_p)$, where ${\bf q}_p = {\bf q}_i + {\bf q}_s $, is the angular spectrum of the pump at the crystal position. In terms of the angular spectrum immediately after the object ${\cal V}_{0}({\bf q}_p)$, which is the Fourier transform of transverse pump field amplitude, we can write

\begin{equation}
\label{eq2}
{\cal V}_{z_2}({\bf q}_p)\propto {\cal V}_0({\bf q}_p)\exp{\left[-i\frac{{q_p}^2}{2k_p}z_1\right]},
\end{equation}
where we have considered the free propagation from the object to the crystal in the paraxial approximation. $k_p$ is the wavenumber of the pump.

The coincidence-counting rate $C({\bbox \rho}_i,{\bbox \rho}_s)$, for detectors located at a distance $z$ from the crystal and transverse coordinates ${\bbox \rho}_i$ and ${\bbox \rho}_s$, is proportional to the fourth-order correlation function $G^{(2,2)}({\bbox\rho}_i, {\bbox\rho}_s)$ given by

\begin{eqnarray}
\label{eq3}
G^{(2,2)}({\bbox\rho}_i, {\bbox \rho}_s) = \langle\psi|\hat E^{(-)}_s({\bbox \rho}_s) \hat E^{(-)}_i({\bbox \rho}_i)\\ \nonumber 
\hat E^{(+)}_i( {\bbox \rho}_i) \hat E^{(+)}_s( {\bbox \rho}_s)|\psi\rangle.
\end{eqnarray}
$\hat E^{(-)}_j( {\bbox \rho}_j)$ and $E^{(+)}_j( {\bbox \rho}_j)$ are the negative and positive frequency parts of the electric field operator which carry information about the effects of the lens and the free propagation from the crystal to the detectors. By using again the paraxial approximation,

\begin{eqnarray}
\label{eq4}
\hat E^{(+)}_j( {\bbox \rho}_j)= \int d {\bf q}_j\int d {\bf q}^\prime_j \hat a( {\bf q}^\prime){\cal T}({\bf q}_j-{\bf q}^\prime_j) \\ \nonumber
\exp \left[ i\left( {\bf q}_j\cdot{\bbox \rho}_j - \frac{q_j^2}{2k_j}(z-z_2)-
\frac{q^{\prime2}_j}{2k_j}z_2\right)\right],
\end{eqnarray}
where $k_j$ is the wave number of the signal ($j=s$) and idler ($j=i$). ${\cal T}({\bf q}_j-{\bf q}^\prime_j)$ is the transmission function of the lens. Assuming that the lens is perfectly transparent and its diameter is much larger than the diameter of the beams, we have

\begin{equation}
\label{eq5}
{\cal T}({\bf q}_j-{\bf q}^\prime_j)=exp{\left(i\frac{f}{2k_j}|{\bf q}_j-{\bf q}^\prime_j|^2\right)},
\end{equation}
where $f$ is the focal length of the lens.

Now, let us calculate the coincidence-counting rate. Eq.\ref{eq3} yields
\begin{eqnarray}
\label{eq6}
 C({\bbox \rho}_i,{\bbox \rho}_s) \propto \biggr |\hat E^{(+)}_i({\bbox \rho}_i) \hat E^{(+)}_s( {\bbox \rho}_s)|\psi\rangle\biggr |^2,
\end{eqnarray}
where

\begin{eqnarray}
\label{eq6b}
\hat E^{(+)}_i \hat E^{(+)}_s|\psi\rangle&=&\int d {\bf q}^\prime_s \int d {\bf q}^\prime_i \int d {\bf q}_s \int d {\bf q}_i  \\ \nonumber
&\times&{\cal V}_{z_2}({\bf q}^\prime_i+{\bf q}^\prime_s){\cal T}({\bf q}_s-{\bf q}^\prime_s){\cal T}({\bf q}_i-{\bf q}^\prime_i)\\ \nonumber
&\times& \exp{\left[ i\left( {\bf q}_s\cdot {\bbox \rho}_s - \frac{q_s^2}{2k_s}(z-z_2)-\frac{q_s^{\prime 2}}{2k_s}z_2\right)\right]}\\ \nonumber
&\times& \exp{\left[ i\left( {\bf q}_i\cdot {\bbox \rho}_i - \frac{q_i^2}{2k_i}(z-z_2)-\frac{q_i^{\prime 2}}{2k_i}z_2\right)\right]}\\ \nonumber
&\times&|0,{\bf q}_s\rangle|0,{\bf q}_s\rangle.
\end{eqnarray}
By using the equations (\ref{eq2}) and (\ref{eq5}) and performing the integrals in ${\bf q}_i$ and ${\bf q}_s$, we get

\begin{eqnarray}
\label{eq8}
 C({\bbox \rho}_i,{\bbox \rho}_s) &\propto& \biggr |\int d {\bf q}^\prime_s \int d {\bf q}^\prime_i  \, {\cal V}_0({\bf q}^\prime_i+{\bf q}^\prime_s) \\ \nonumber
&\times&\exp{\left[-i\frac{|{\bf q}^\prime_i+{\bf q}^\prime_s|^2}{2k_p}z_1\right]} \\ \nonumber
&\times&\exp{\left[-i\frac{q_s^{\prime 2}}{2k_s}\left(z_2-f+\frac{f^2}{z-z_2-f}\right)\right]}\\ \nonumber  
&\times&\exp{\left[-i\frac{f}{z-z_2-f}{\bf q}_s^{\prime}\cdot {\bbox \rho}_s\right]}  \\ \nonumber
&\times&\exp{\left[-i\frac{q_i^{\prime 2}}{2k_i}\left(z_2-f+\frac{f^2}{z-z_2-f}\right)\right]}\\ \nonumber  
&\times&\exp{\left[-i\frac{f}{z-z_2-f}{\bf q}_i^{\prime}\cdot {\bbox \rho}_i\right]} 
\biggr |^2.
\end{eqnarray}
The lens in Fig. 1 is chosen such that $z_2+z_1$ is the distance object-lens $O$ and $z-z_2$ is the distance lens-image $I$. By using the thin-lens equation
$1/f=1/I+1/O$, eq.(\ref{eq8}) can be written as

\begin{eqnarray}
\label{eq9}
 C({\bbox \rho}_i,{\bbox \rho}_s) &\propto& \biggr |\int d {\bf q}^\prime_s \int d {\bf q}^\prime_i  \, {\cal V}_0({\bf q}^\prime_i+{\bf q}^\prime_s) \\ \nonumber
&\times&\exp{\left[i\frac{|{\bf q}^\prime_i-{\bf q}^\prime_s|^2}{2k_p}z_1\right]} \\ \nonumber
&\times&\exp{\left[-i\frac{O}{I}({\bf q}_s^{\prime}\cdot {\bbox \rho}_s+{\bf q}_i^{\prime}\cdot {\bbox \rho}_i)\right]}
\biggr |^2,
\end{eqnarray}
where we have considered the degenerated case, with $k_i=k_s=k_p/2$.

Introducing relatives variables

\begin{eqnarray}
\label{eq10}
{\bf q}^\prime_\pm={\bf q}^\prime_i \pm {\bf q}^\prime_s \\ \nonumber
{\bbox \rho}_\pm= {\bbox \rho}_s \pm {\bbox \rho}_s
\end{eqnarray}
eq.(\ref{eq9}) can be rewritten as

\begin{eqnarray}
\label{eq11}
C({\bbox \rho}_i,{\bbox \rho}_s) &\propto& \biggr |\int d {\bf q}^\prime_+{\cal V}_0({\bf q}^\prime_+) \exp{\left[-i\frac{O}{I}{\bf q}_+^{\prime}\cdot{\bbox \rho}_+\right]}\\ \nonumber
&\times&\int d {\bf q}^\prime_- \exp{\left[-i\frac{O}{2I}{\bf q}_-^{\prime}\cdot{\bbox \rho}_-\right]}\exp{\left[i\frac{q^{\prime 2}_-}{2k_p}z_1\right]}
\biggr |^2.
\end{eqnarray}
The first integral is the Fourier transform of the angular spectrum of the pump immediately after the object, while the second one is given by

\begin{eqnarray}
\label{12}
\int d {\bf q}^\prime_- \exp{\left[-i\frac{O}{I}{\bf q}_-^{\prime}\cdot{\bbox \rho}_-\right]}\exp{\left[i\frac{q^{\prime 2}_-}{2k_p}z_1\right]}=\\ \nonumber
\exp{\left[ -i\rho_-^2\frac{O^2k_p}{2I^2z_1}\right]},
\end{eqnarray}
which is a phase factor, whose square modulus is equal to one.
Therefore, the coincidence-counting rate at the detection plane is

\begin{equation}
\label{eq13}
C({\bbox \rho}_i,{\bbox \rho}_s) \propto \biggr | W_0\left( \frac{O}{I}{\bbox \rho}_+\right) \biggr |^2.
\end{equation}
This result shows that the coincidence-counting rate is proportional to the transverse intensity profile of the pump immediately after the object $\cal O$ as a function of the transverse relative coordinate of the signal and idler detectors, with a scale factor that takes into account the magnification factor $m=I/O$. Therefore, by placing an object in the pump, before the crystal, and a lens in the twin beams, we obtain the image of the object in a given image plane by detecting the twin beams in coincidence, with a magnification factor $m$, that depends on the distance object-lens and lens-image. 

To compare the result for the coincidence-counting rate with the intensity of the pump we have calculated the intensity of the pump detected by D3. The intensity of the pump $|W({\bbox \rho})|^2$ is given by

\begin{equation}
\label{eq13b}
|W({\bbox \rho})|^2 \propto \biggr | W_0\left( \frac{O}{I}{\bbox \rho}\right) \biggr |^2,
\end{equation}
which is the image of the object ${\cal O}$ with a magnification factor $m=I/O$. Comparing this result with eq. (\ref{eq13}), we see that by displacing detectors D1 and D2 simultaneously, such that each one is looking at the same point along the transverse direction, the image formation in coincidence by placing the object in the pump and the lens in the down-converted beams is analogous to putting the object and the lens in the pump and measuring its intensity.

\section{Experiment and results}

\begin{figure}[h]
\includegraphics*[width=8cm]{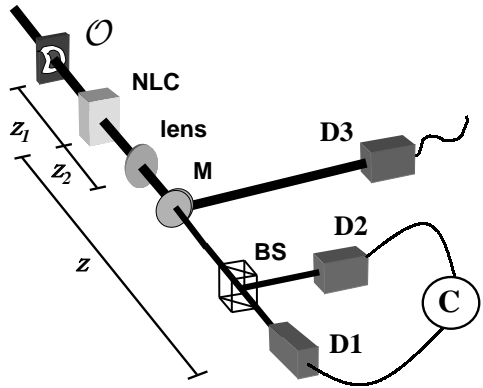}
\begin{center}
\caption{Experimental setup.}
\end{center}
\label{fig1}
\end{figure}

The experiment has been performed in one dimension, by using a double-slit as the object ${\cal O}$. The experimental setup is shown in Fig. \ref{fig1}. A cw He-Cd laser operating at 442 nm pumps a 1-cm-long {\em Lithium Iodate} (LiIO$_3$) crystal cut for type-I phase
matching. The polarization of the pump and the optical axes of the crystal are aligned such that signal and idler emerge from the crystal collinearly with the pump, at the degenerate wavelength 884 nm. M is an optical window that reflects the pump and transmits the twin beams. After that, the twin beams are split in a beam-splitter and detected with vertically oriented thin slits of about 0.2 mm, 12.5-mm focal-length lenses, and 10 nm bandwidth interference filters centered at 884 nm at the entrance of the single-photon counting modules (D1 and D2). The same detection assembly with proper optical filters (D3) is used to detect the pump. Single and coincidence counts are registered by scanning the detectors horizontally. 

\begin{figure}[h]
\begin{center}
\includegraphics*[width=6cm]{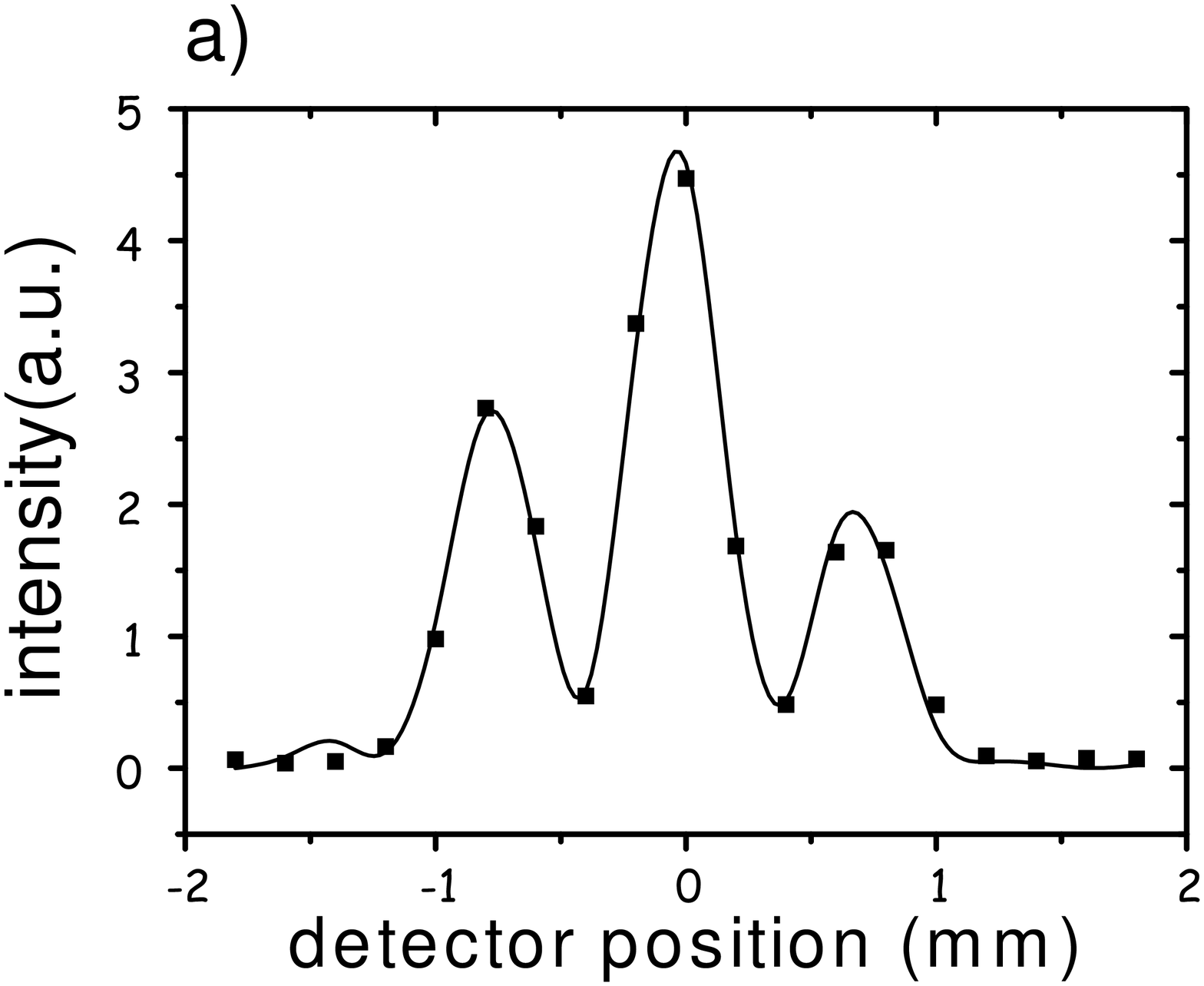}
\includegraphics*[width=6cm]{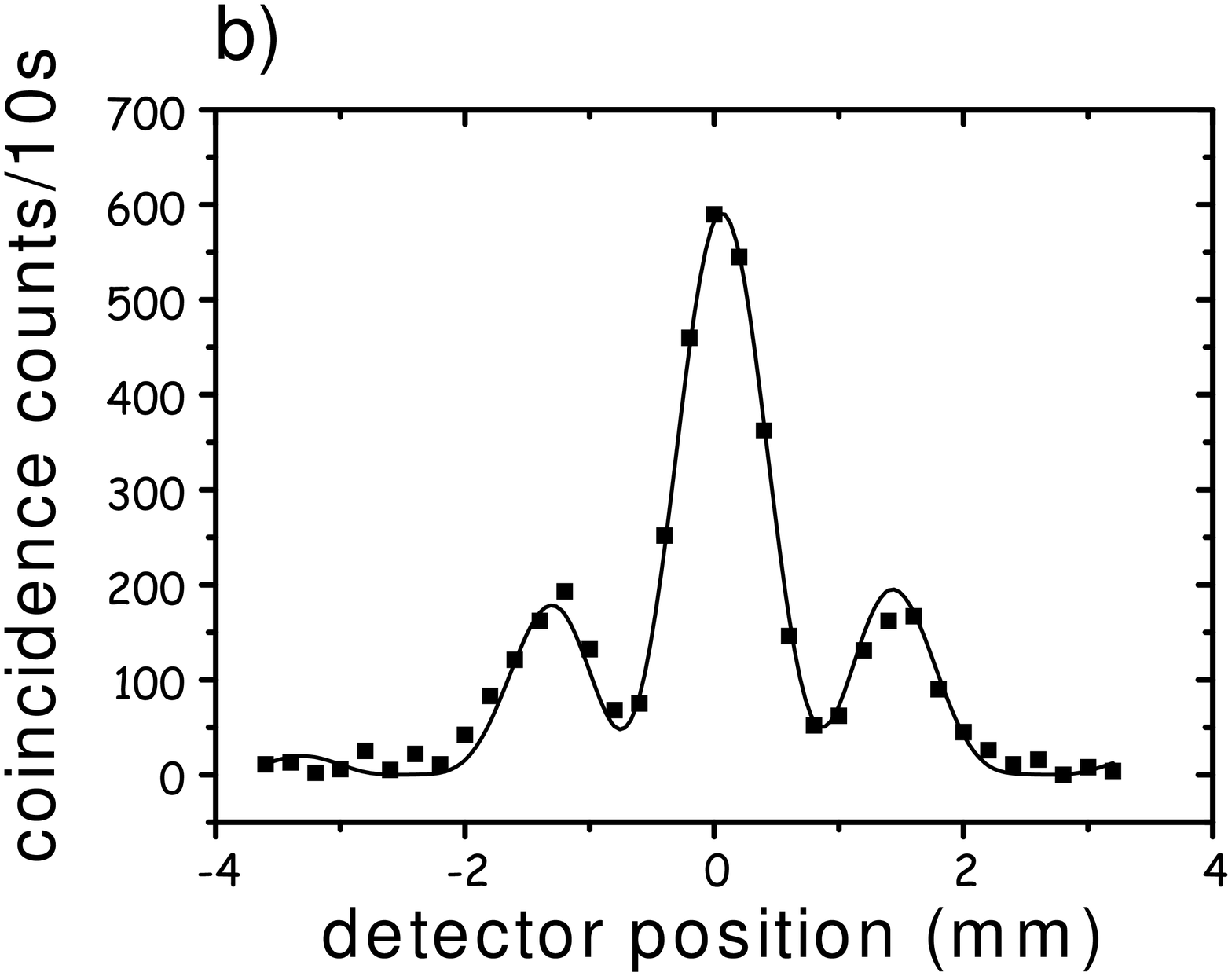}
\includegraphics*[width=6cm]{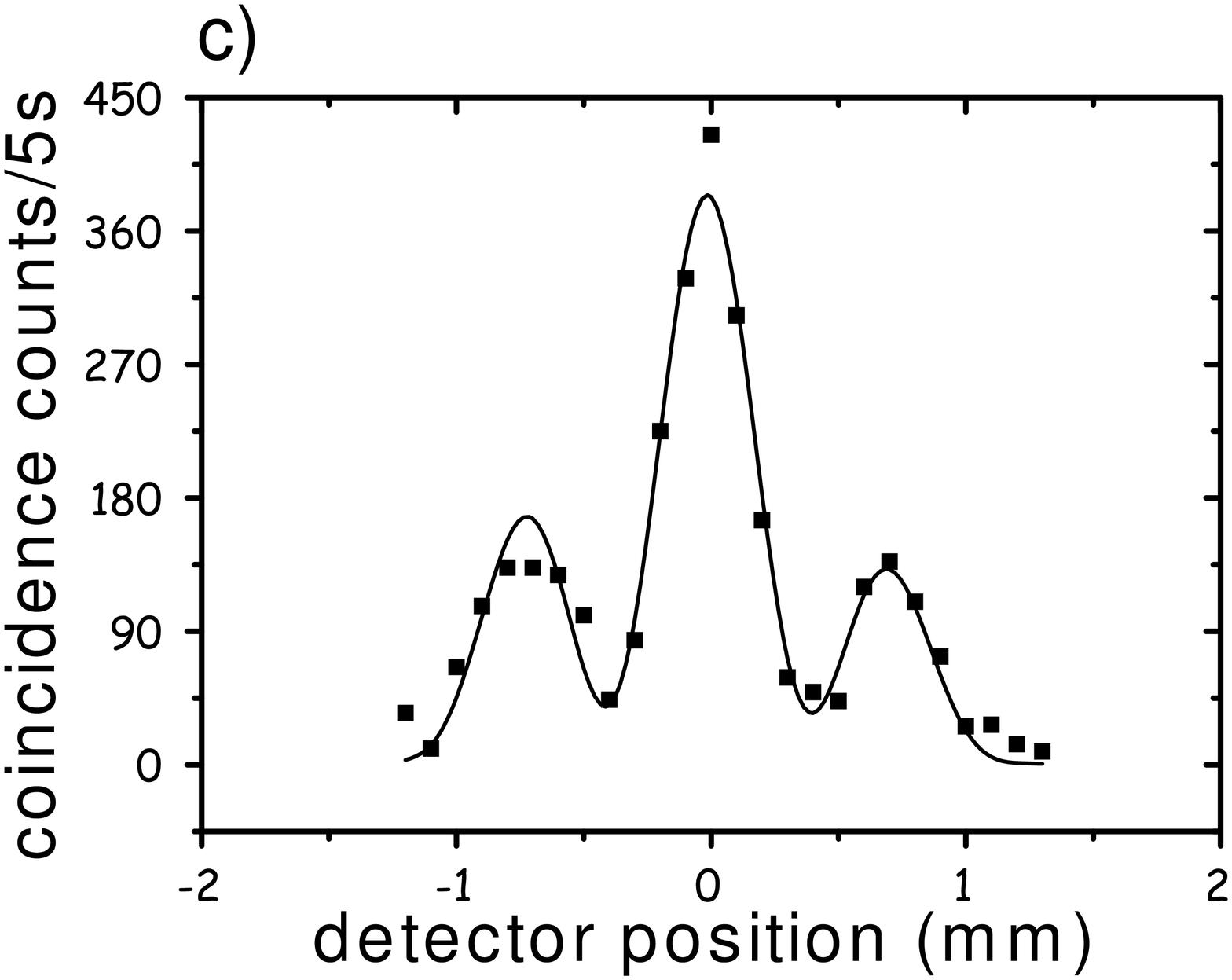}
\end{center}
\caption{(a) Pump beam intensity; (b) coincidence counts measured by scanning $D1$ and keeping $D_2$ detector fixed; (c) coincidence counts measured by scanning D1 and D2 simultaneously.}
\label{fig2}
\end{figure}

First, we have observed the angular spectrum transfer from the pump to the twin beams by inserting a double slit in the pump about 34 cm from the crystal. All detectors are placed about 70 cm from the crystal. In Fig. \ref{fig2}a the interference pattern observed in the pump beam scanning D3 is shown. Scanning D1 and keeping D2 fixed, the result for the coincidence counts is shown in Fig. \ref{fig2}b, exhibiting the transferred interference pattern. We have also performed this measurement scanning D1 and D2 simultaneously and in the same sense. The result is shown in Fig. \ref{fig2}c, also exhibiting an interference pattern, but now with the same spatial frequency of the pattern in the pump.

\begin{figure}[h]
\begin{center}
\includegraphics*[width=6cm]{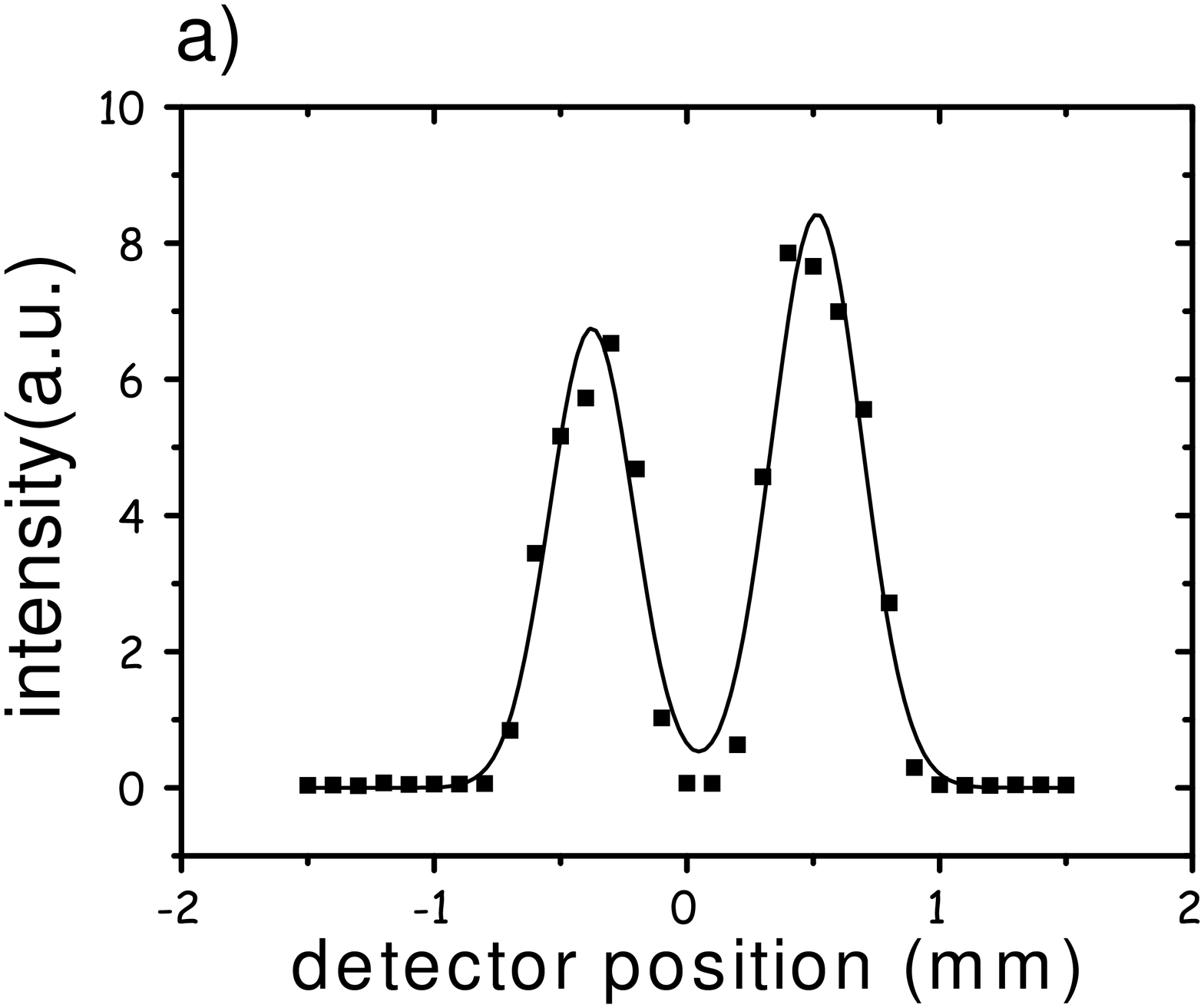}
\includegraphics*[width=6cm]{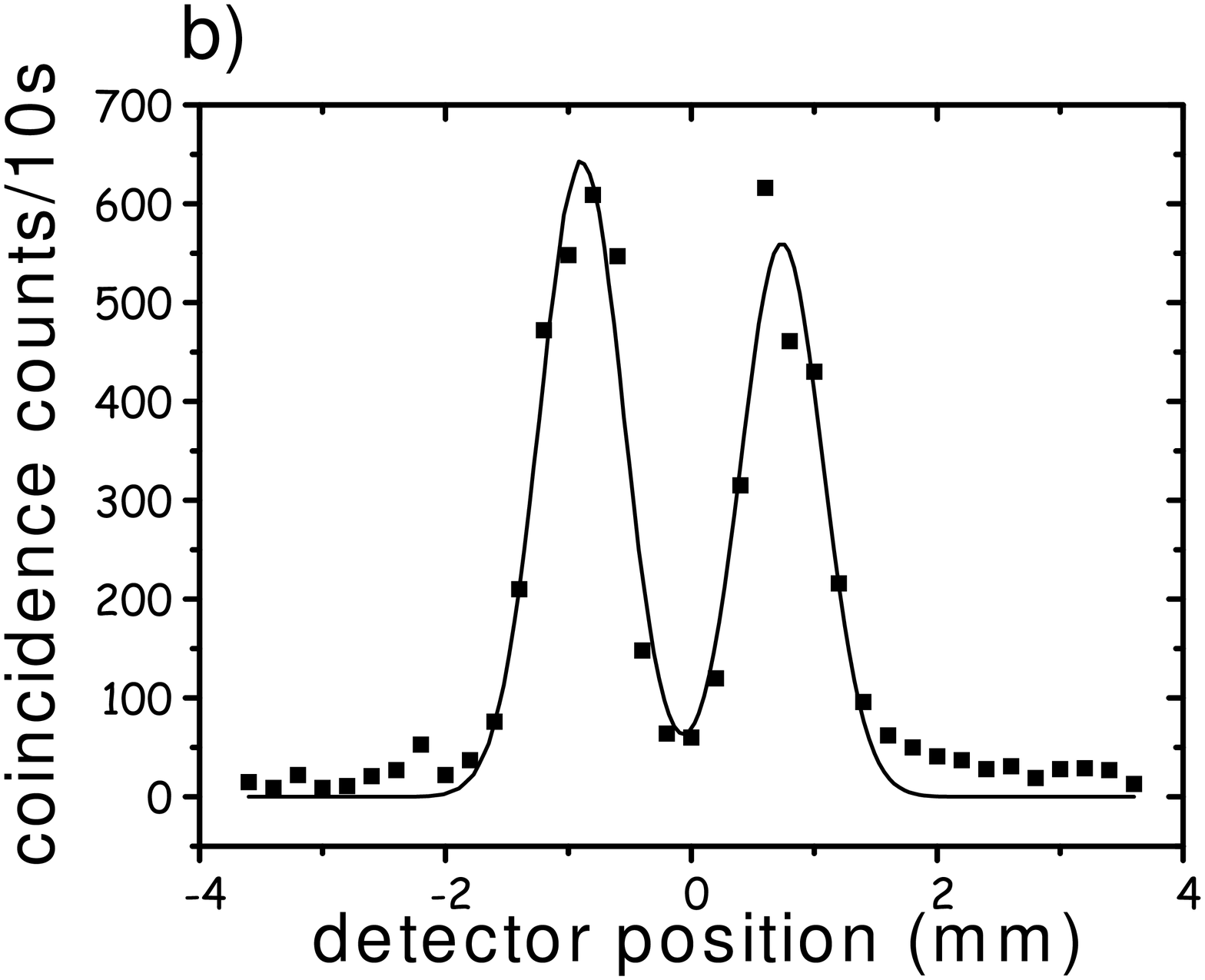}
\includegraphics*[width=6cm]{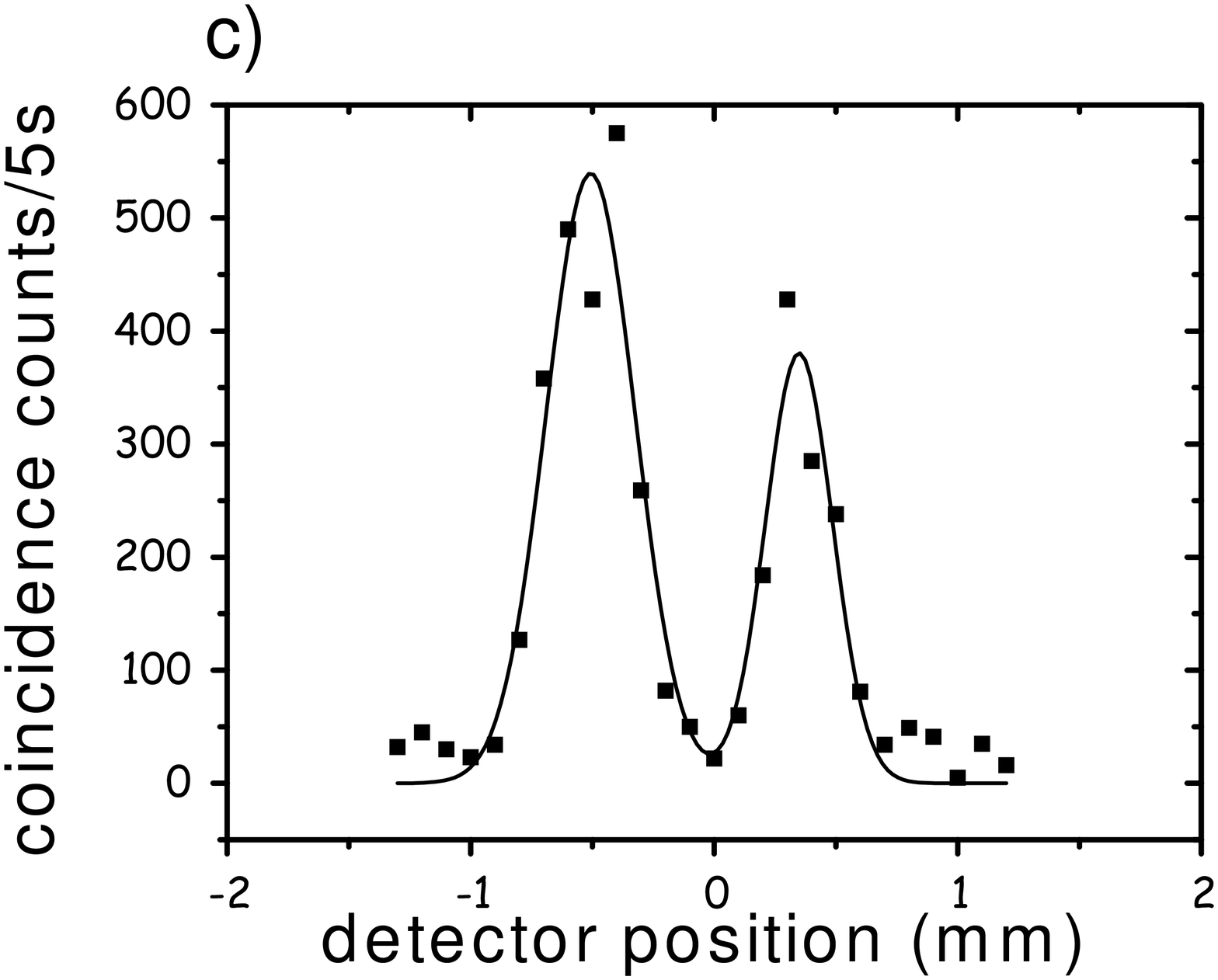}
\end{center}
\caption{(a) Pump beam intensity; (b) coincidence counts measured by scanning $D1$ and keeping $D_2$ detector fixed; (c) coincidence counts measured by scanning D1 and D2 simultaneously.}
\label{fig3}
\end{figure}

Second, we have inserted a 25 cm focal length lens about 7 cm from the crystal, such that the image of the double-slit is formed at the detection plane. In Fig. \ref{fig3}a
the intensity profile of the pump is shown, where we can see a two peaks intensity pattern corresponding to the image of the double-slit. Scanning D1 and keeping D2 fixed, we observe the image of the double-slit in the coincidence counts. This result is shown in Fig. \ref{fig3}b. Comparing Fig. \ref{fig3}a with Fig. \ref{fig3}b, the image of the double-slit in coincidence is two times larger than the image in the pump. This is due to the fact that when one of the detectors is kept fixed we have ${\bbox \rho}_s=0$, for example, and the coincidence-counting rate becomes
\begin{equation}
\label{eq14}
C({\bbox \rho}_i,{\bbox \rho}_s) \propto \biggr | W_0\left( \frac{1}{2m}{\bbox \rho}_i\right) \biggr |^2,
\end{equation}
where we have a 2 factor that doubles the size of the image. The magnification factor of $m\approx $1.5 is the same for all images.

Scanning both D1 and D2 simultaneously and in the same sense, the coincidence counts show the image of the double-slit with the same scale of Fig. \ref{fig3}a. This result is shown in Fig. \ref{fig3}c. In this case we have set ${\bbox \rho}_i={\bbox \rho}_s={\bbox \rho}$, and the coincidence-counting rate becomes

\begin{equation}
\label{eq15}
C({\bbox \rho}_i,{\bbox \rho}_s) \propto \biggr | W_0\left( \frac{1}{m}{\bbox \rho}\right) \biggr |^2,
\end{equation}
which is analogous to the measurement of the intensity of the pump showed in Fig. \ref{fig3}a.

\section{conclusion}

In conclusion, we have demonstrated theoretical and experimentally, that the image of a given object placed in the pump can be formed in the twin beams by manipulating the transferred angular spectrum and detecting coincidence. Even though we have analyzed the collinear and degenerated case, our results can be easily generalized to the noncollinear and nondegenerated case.

\begin{acknowledgments}

The authors acknowledge financial support from the Brazilian 
agencies CNPq, CAPES, PRONEX, Institutos do Mil\^enio-Informa\c{c}\~ao Qu\^antica, FAPERJ and FUJB.

\end{acknowledgments}

\end{document}